\documentclass[fleqn,usenatbib]{mnras}
\usepackage{newtxtext,newtxmath}
\usepackage[T1]{fontenc}

\usepackage{amsmath}
\usepackage{amsfonts}
\usepackage{graphicx}
\usepackage{xcolor}

\title[Constraining $H_0$ Via Extragalactic Parallax]{Constraining $H_0$ Via Extragalactic Parallax}
\author[Ferree and Bunn]{
Nicolas C. Ferree,$^{1}$
Emory F. Bunn$^{1}$\thanks{E-mail: ebunn@richmond.edu}
\\
$^{1}$Physics Department, University of Richmond, Richmond, VA  23173, USA\\}

\date{Accepted XXX. Received YYY; in original form ZZZ}

\pubyear{2021}

\begin{document}
\label{firstpage}
\pagerange{\pageref{firstpage}--\pageref{lastpage}}
\maketitle

\begin{abstract}
We examine the prospects for measurement of the Hubble parameter $H_0$  via observation of the secular parallax of other galaxies due to our own motion relative to the cosmic microwave background rest frame. Peculiar velocities make distance measurements to individual galaxies highly uncertain, but a survey sampling many galaxies can still yield a precise $H_0$ measurement.
 We use both a Fisher information formalism and simulations to forecast errors in $H_0$ from such surveys, marginalizing over the unknown peculiar velocities. The optimum survey observes $\sim 10^2$ galaxies within a redshift $z_\mathrm{max}=0.06$. The required errors on proper motion are comparable to those that can be achieved by Gaia and future astrometric instruments.
A measurement of $H_0$ via parallax has the potential to shed light on the tension between different measurements of $H_0$. 
\end{abstract}


\bibliographystyle{mnras}

\section{Introduction}
Parallax is a common tool to measure distances to relatively nearby astrophysical objects. Parallax distances to nearby stars form the first rung of the ``distance ladder'' \citep[e.g.,][and references therein]{gaiastellar}.

Because they depend only on geometry, parallax distances are particularly robust. As a result, it would be extremely desirable to use the method on cosmological scales, bypassing the other steps in the ladder. Traditional parallax, using the Earth's orbit as a baseline, results in unmeasurably small parallax angles for sources at cosmological distances. \citet{kardashev} noted that a larger signal could be obtained by using the solar system's motion relative to the cosmological rest frame to produce a longer baseline. From the cosmic microwave background dipole, we infer that our velocity relative to this frame is $78\,\mathrm{AU\,yr}^{-1}$ \citep{hinshaw}, leading to a secular proper motion $(78\mu\mathrm{as\,yr}^{-1})D_\mathrm{Mpc}^{-1}\sin\beta$ for a source located at an angular diameter distance\footnote{We will assume a cosmologically flat Universe throughout this paper. In a spatially curved universe, there is a distinction between the parallax distance and the angular diameter distance \citep[][and references therein]{hoggdistance}.
Because we focus on sources at low redshift, any corrections due to spatial curvature would be extremely small and would not affect our conclusions.}
$D_\mathrm{Mpc}$ megaparsecs in a direction making an angle $\beta$ with the known CMB dipole direction. 

For the near future, the uncertainty in a single parallax measurement is too high to be of use in determination of distances on a cosmological scale; however, advances in telescope technology provide the possibility of using cosmological parallax in a statistical analysis \citep{Gaia_parallax}. Here, we analyze the usage of parallax measurements to constrain the Hubble parameter $H_0$. Should this method be feasible in the near future, it would provide an independent determination of $H_0$ that might help to resolve the current ambiguity in its value \citep{planck, riess_hubble}.

In this paper, we will explore the space of design parameters for a survey of galaxy parallax measurements. We use both a Fisher formalism and simulations to determine the error in a measurement of $H_0$ from a given survey, taking into account both proper motion measurement errors and the peculiar velocities of the source galaxies.
For redshift-limited surveys with varying numbers of galaxies and redshift limits, we determine the required precision with which proper motions need to be measured in order to yield an $H_0$ measurement with a given error. 

Our purpose in this paper is to perform a preliminary study of the feasibility of an observation program of this sort. The calculations are based on approximations, in particular regarding the peculiar velocity field, that will have to be replaced with more sophisticated methods if and when such a survey is actually performed. We will argue below that our approach is adequate to give an initial estimate of the required errors and of the optimum survey design. The error forecasts we present are, of course, not as precise as those that would arise from a detailed simulation of a specific survey design. We regard such a detailed calculation as premature at this stage, although it will certainly be required at a later stage if attempts to measure $H_0$ via this method are actually made.

Some previous explorations of this possibility have focused on observations of quasars \citep{dingcroft,bachchan}. Other work, like ours, has focused on closer sources for which the predicted signal is larger \citep{croft,Gaia_parallax,hall}. \citet{hall} develop a harmonic-space formalism for modeling the galaxy peculiar velocities, whereas we have chosen to work in real space. \citet{croft} and \citet{Gaia_parallax} consider data sets modeled on particular present or future observing campaigns, whereas we have chosen to adopt an exploratory approach, asking what observing strategy would be optimal for this particular measurement.

We focus specifically on measuring $H_0$, but extragalactic proper motion surveys can be used to study a number of additional interesting questions \citep[e.g.,][and references therein]{darling2}. Of particular note is the ability to measure the \textit{acceleration} of our solar system's motion via secular aberration drift \citep{truebenbacksad}. Probing the connection between the galaxy proper motion field and the matter power spectrum is also potentially of great interest \citep{darling}.

The remainder of this paper is structured as follows. Section 2 contains the details of our methods, including the computation of marginalized likelihoods needed for both Fisher calculations and maximum-likelihood estimation in our simulations, along with numerous other details. Section 3 contains a summary of our results, and a discussion is found in Section 4.

\section{Methods}

\subsection{Assumptions}
We wish to measure $H_0$ using parallax. Suppose that we have a sample of galaxies, each of which has a measured angular position, redshift, and proper motion. 

From redshift alone, we are only able to determine a combination of $H_0$ and the distance to a galaxy (with some uncertainty due to the galaxy's unknown radial peculiar velocity). We then use the proper motion of the galaxy to constrain the distance alone, allowing us to separate the distance from $H_0$.

We begin with some simplifying assumptions. First, we assume that the error in angular position measurements is negligible compared to other errors, so we treat the angular position of each galaxy as perfectly known. Second, we assume that the velocity of the solar system with respect to the cosmic microwave background has been measured with a high degree of accuracy, so we treat it too as being perfectly known. Third, we assume that galaxies' peculiar velocities are uncorrelated and that each component of the peculiar velocity is drawn from a normal distribution with mean zero and standard deviation $\sigma_v$. While actual galaxy peculiar velocities are correlated, we will show in Section \ref{sec:correlation_results} that the error introduced by this approximation is acceptable. Fourth and finally, we assume that all noise is normally distributed with mean zero.

\subsection{Likelihood Function and Priors}
We place the Earth at the center of our coordinate system and suppose that our survey has $N$ galaxies. Since the angular position of each galaxy is assumed to be known perfectly, we may associate with each galaxy $j$ an orthonormal basis of vectors $\hat{R}_j, \hat{\theta}_j, \hat{\phi}_j$. We denote the peculiar velocity of the galaxy as $\vec{v}_j$, the peculiar velocity of the solar system as $\vec{v}_0$, and the relative peculiar velocity of the galaxy as $\vec{V}_j = \vec{v}_j - \vec{v}_0$. Therefore $V_{jr} = \vec{V_j} \cdot \hat{R}_j$, $V_{j\phi} = \vec{V}_j \cdot \hat{\phi}_j$, and $V_{j\theta} = \vec{V_j} \cdot \hat{\theta}_j$. We denote the angular diameter distance to the $j^{\text{th}}$ galaxy as $R_{jA}$ and the comoving distance to the $j^{\text{th}}$ galaxy as $R_{j}$. Finally, we let $Z(H_0,R_{j})$ denote the cosmological redshift (i.e., the redshift in the absence of peculiar velocity) of a galaxy at distance $R_{j}$ for a given value of $H_0$.

We may now write a function $\lambda = - 2 \text{ln}(L) = - 2 \text{ln}(\mathcal{L})+\mbox{const}$, where $\mathcal{L}$ is the likelihood function for our model and $L$ is an unnormalized function proportional to the likelihood:
\begin{align}
  \lambda &= 
  \sum_{j=1}^N
  \left(
  \frac{
    \left[z_j
      - \frac{c Z(H_0,R_{j}) + V_{jr}}{c}\right]^2
  }{\sigma_{zj}^2} +
  \frac{\left[\dot{\phi_j} - \frac{V_{j\phi}}{R_{jA}
        \text{sin}(\theta_j) }\right]^2}{\sigma_{\dot{\phi_j}}^2}
\right.
  \nonumber\\
  &\hskip 1.6in\left.
  +  \frac{\left[\dot{\theta_j} -
      \frac{V_{j\theta}}{R_{jA} }\right]^2}{\sigma_{\dot{\theta} j}^2} \right)
\end{align}
As usual, the likelihood is the probability density of the observed data, given a set of theoretical parameters. In the above expression, the observed data are $z_j$ (the redshift) and $\dot{\phi_j}$ and $\dot{\theta_j}$ (which describe the proper motion). The theoretical parameters are the Hubble parameter $H_0$, the comoving distances $\{ R_j \}$, and the peculiar velocities $\{ \vec{v}_j \}$. (The angular diameter distances $ \{ R_{jA} \}$ are determined by $ \{ R_{j} \}$ and $H_0$  so are not independent parameters.) The other quantities appearing in this expression, specifically 
the measurement uncertainties $\sigma_{\dot\theta_j},\sigma_{\dot\phi_j},\sigma_{z_j}$, are assumed to be known. Recall that $\theta_j$ and $\phi_j$, measurements of the angular position, are assumed to be known perfectly. We therefore treat all of these values as constants. This leaves $H_0, \{ R_j \}, $ and $ \{ \vec{v}_j \}$ as the parameters in our model.

Our model considers all of the galaxies to be independent of each other, so the likelihood function for a given galaxy and set of parameters depends only on the measurements for that galaxy. Thus,

\begin{equation}
    L(H_0, \{ R_j \}, \{ \vec{v}_j \}) = \displaystyle \prod_{j=1}^N L_j(H_0, R_j, \vec{v}_j),
\end{equation}
where $L_j = e^{- \frac{1}{2} \lambda_j}$ and $\lambda_j$ is the $j^{\text{th}}$ term in the sum $\lambda$.

To determine our prior, we assume that the proper volume density of galaxies is uniform and that peculiar velocities have independent components drawn from a normal distribution with mean zero and standard deviation $\sigma_v$.

Since we assume that galaxies are drawn at random within our survey volume, the prior $P_j(H_0,R_{j})$ on $H_0$ and $R_{j}$ satisfies $dP_j \propto n R_{j}^2 dR_{j}$, where $n$ is the number density of galaxies in redshift space. This density is proportional to $H_0^3$ in real space, so  $dP \propto H_0^3 R_{j}^2 dR_{j}$. Our model assumes that all of the galaxies are independent of each other, so the prior on $H_0$ and $ \{ R_{j} \}$ is the product of the individual priors: 
\begin{equation}
    dP(H_0, \{ R_j \}) = \prod_{j=1}^N dP_j(H_0, R_{j}) \propto \prod_{j=1}^N H_0^3 R_{j}^2 dR_{j}.
\end{equation}

Furthermore, our assumptions about the distribution of peculiar velocity components lead to a peculiar velocity prior 
\begin{equation}
P(\{ \vec{v}_k \}) = \prod_{j=1}^N P_j(\vec{v}_j) = \prod_{j=1}^N e^{\frac{-||\vec{v}_j||^2}{2 \sigma_v^2}}.  
\end{equation}
(See Section \ref{sec:correlation_results} for further discussion of 
galaxy peculiar velocities.)
Therefore the unnormalized posterior probability distribution for the survey becomes
\begin{equation}
    A(H_0, \{ R_j \}, \{ \vec{v}_j \}) \propto \displaystyle \prod_{j=1}^N H_0^3 R_j^2  e^{\frac{-||\vec{v}_j||^2}{2 \sigma_v^2}} L_j(H_0, R_j, \vec{v_j}).
\end{equation}

The peculiar velocities $\{ \vec{v}_j \}$ are nuisance parameters, so we perform an analytic marginalization. The marginalized likelihood is $\prod_jB_j$, where
\begin{equation}
B_j(R_j, H_0) \propto H_0^3 R_j^2 \displaystyle \iiint \limits_{-\infty}^{+\infty} e^{\frac{-||\vec{v}_j||^2}{2 \sigma_v^2}} L_j \,dv_{jr}\, dv_{j\phi}\, dv_{j\theta}.
\end{equation}
This yields 
\begin{equation}
    B_j(R_j, H_0) \propto H_0^3 R_j^2  e^{-\frac{1}{2}  \beta_j},
\end{equation}
where
\begin{align}
\beta_j(R_j, H_0) &=  -  \text{ln} \left (\frac{R_{jA}^4} {(\sigma_v^2 + R_{jA}^2 \sigma_{\dot{\theta} j } ^2) (\sigma_v^2 \text{csc}^2(\theta_j) + R_{jA}^2 \sigma_{\dot{\phi} j } ^2)} \right )\nonumber\\
&\quad + \frac{ (-cZ(H_0,R_{j}) + v_{0r} + c z_j) ^2 }{\sigma_v^2 + c^2 \sigma_{zj}^2}     
+ \frac{ (\dot{\theta_j} R_{jA} + v_{0 \theta})^2}{\sigma_v^2 + R_{jA}^2 \sigma_{\dot{\theta} j }^2} \nonumber\\
&\quad + \frac{ (\dot{\phi_{j} }R_{jA} + \text{csc}(\theta_j) v_{0 \theta})^2  }{ \text{csc}^2(\theta_j) \sigma_v^2 + R_{jA}^2 \sigma_{\dot{\phi_j}}^2} .
\end{align}

The distances $\{ R_{j} \}$ are also nuisance parameters, so we marginalize over them as well. This integral cannot be done analytically, so we resort to numerical methods. Fortunately, the independence of the galaxies means that we can write the $N$-dimensional integral as a product of $N$ one-dimensional integrals. Thus, our final expressions for the posterior probabilities $\{ \mathcal{L}_j \}$ have only $H_0$ as a parameter:
\begin{equation}
\mathcal{L}_j(H_0) \propto \int \limits_{0} ^ {\infty} B_j(R_j,H_0)dR_j
\label{l2}
\end{equation}

Therefore
\begin{equation}
    \mathcal{L}(H_0) \propto \displaystyle \prod_{j=1}^N \mathcal{L}_j(H_0),
\end{equation}
so we now have an expression for our posterior probability distribution $\mathcal{L}$ marginalized over all nuisance parameters.

\subsection{Fisher Information}
\label{sec:fisher}
Recall that for a single parameter $H_0$ and posterior probability distribution $\mathcal{L}(H_0)$, the Fisher Information $F(H_0)$ is given by 
\begin{equation}
    F(H_0) = \left \langle - \frac{d^2 \text{ln} ( \mathcal{L}(H_0)) }{dH_0^2} \right \rangle.
\end{equation}
Since the measurements for different galaxies are independent, we know that the Fisher information for a sample of $N$ galaxies is $F_N(H_0) = N  F_j(H_0)$, where $F_j (H_0)$ is the Fisher information for a single galaxy. Thus, we may concern ourselves with calculating only $F_j (H_0)$, which is
\begin{equation}
    F_j(H_0) = \left \langle - \frac{d^2 \text{ln} (\mathcal{L}_j(H_0)) }{dH_0^2} \right \rangle.
\end{equation}
We therefore wish to compute 
\begin{equation}
    \left \langle - \frac{d^2 \text{ln} ( \mathcal{L}_j (H_0))}{dH_0^2} \right \rangle = \left \langle - \frac{ \mathcal{L}_j(H_0) \frac{d^2 \mathcal{L}_j}{dH_0^2} - \left(\frac{d \mathcal{L}_j}{dH_0}\right)^2 }{\mathcal{L}_j(H_0)^2} \right \rangle.
\end{equation}
 We compute this value via Monte-Carlo integration: for a fixed set of experimental parameters, we randomly generate many sets of data (with a uniform proper volume density) and find $- \frac{d^2 \text{ln} (\mathcal{L}_j (H_0)) }{dH_0^2}$ for each, taking the mean as the expected value. This yields the Fisher information for a single galaxy, which we then multiply by $N$ to find the Fisher information for a survey of $N$ galaxies. To simplify this calculation, we use the low-redshift approximation $cZ(H_0,R_{j}) = H_0 R_{j}$.
 
 Given the expected value of the Fisher information, we then compute a lower bound on $\sigma_{H_0}$, the standard deviation of the maximum-likelihood estimator of $H_0$, via the Cram\'er-Rao Inequality:
 \begin{equation}
     \sigma_{H_0} \geq \frac{1}{\sqrt{F_N (H_0)}}.
     \label{eq:cramer-rao}
 \end{equation}
This inequality applies to any unbiased estimator of $H_0$. In Section \ref{sec:fisher_results}, we will compare it to the maximum-likelihood estimator. 

While the maximum-likelihood estimator is biased (as we will see in Section \ref{sec:fisher_results}), the bias is typically small. Furthermore, the correct expression for the Cram\'er-Rao inequality for a biased estimator $\hat{\theta}$ of a parameter $\theta$ is
\begin{equation}
    \text{var}(\hat{\theta}) \geq \frac{(1+b'(\theta))^2}{F(\theta)},
\end{equation}
where $b(\theta)$ is the bias and $F(\theta)$ is the Fisher information \citep{jaynes}. In the case where $b'(\theta)$ is small, this lower bound is well-approximated by $1/F(\theta)$. We will show in Section \ref{sec:fisher_results} that the maximum-likelihood estimator of $H_0$ is in such a regime, and therefore equation (\ref{eq:cramer-rao}) is approximately true.

\subsection{Simulation}
\label{sec:simulation}
We also compute the uncertainty in $H_0$ via simulation. The simulation has the advantage of allowing all redshifts and angular diameter distances to be calculated from a numerical solution of the Friedmann equation, so the low-redshift approximation does not enter into the simulation.

Given the maximum redshift of the survey, we use an interpolation of our numerical solution of the Friedmann equation to find the comoving distance that corresponds to a cosmological redshift equal to the maximum survey redshift. We then randomly generate galaxies with a uniform proper volume density throughout a sphere whose radius is twice that of the calculated comoving distance. Each galaxy is then given a random peculiar velocity, with each component drawn from a normal distribution with mean zero and standard deviation $\sigma_v$. Galaxies whose resulting redshifts exceed the redshift limit of the survey are then discarded. This process is iterated until the desired number of galaxy observations have been generated. We then assume that the two components of each galaxy's proper motion have been measured with some uncertainty $\sigma_p$. We generate the observed data by adding noise with this standard deviation to the ``true'' proper motion components corresponding to the given distance and peculiar velocity.

Then, given a set of data, we numerically optimize $\mathcal{L}$ to find the maximum-likelihood estimate of $H_0$. We repeat this for many randomly generated surveys in order to compute the expected value and standard deviation of the maximum-likelihood estimator of $H_0$. We take the standard deviation of the estimator to be the uncertainty in the measurement of $H_0$.

\subsection{Velocity Field Reconstruction}
\label{sec:velocityreconstruction}
The procedure described above assumes that galaxy peculiar velocities are \textit{uncorrelated} and \textit{unknown}. As we will describe below, we consider hypothetical surveys in which galaxies are spaced farther apart than the velocity correlation length, to reduce the effect of correlated peculiar velocities. In order to further assess the effect of these assumptions, we perform simulations in a subset of our parameter space for models in which galaxy peculiar velocities are correlated but reconstructed with some uncertainty.

The likelihood function and structure of the simulation remains essentially the same. However, in our \emph{primary tests}, when we generate the peculiar velocities of the galaxies, we consider two parts: a correlated component due to a velocity field that is a realization of a Gaussian random process with a given coherence length, and an uncorrelated component where each component is drawn independently from a normal distribution. We further imagine that the correlated component is known and the uncorrelated component is unknown. These assumptions are meant to approximate the idea that reconstructions of peculiar velocities will do much better at determining the large-scale coherent component of the velocity field, so the unknown residual will be much less correlated. 

Given the importance of velocity correlations in the actual Universe, we conducted secondary tests to verify that our treatment of velocity correlations is reasonable. The worst-case scenario for our analysis is an experiment in which the actual peculiar velocities are correlated but the analysis in the future experiment treats them as uncorrelated. We analyze this case by generating a peculiar velocity field that is a realization of a Gaussian random process with a coherence length of $50$ Mpc and using this field to generate data as described in Section \ref{sec:simulation}. We then fit $H_0$ to this data exactly as described in Section \ref{sec:simulation}, as if the simulated peculiar velocities had no correlations.

\subsection{Extragalactic Statistical Parallax}
In most cases of interest, the most useful proper motion signal comes from the motion of the Earth with respect to the CMB, since we consider this to be perfectly known. However, there is in principle another effect in the signal, which we will describe as extragalactic statistical parallax. This effect is due to the peculiar velocities of the observed galaxies; while each peculiar velocity is unknown, we have assumed that we know the distribution they are drawn from. This knowledge of a typical peculiar velocity lets us estimate a typical distance for a galaxy (given its proper motion). With this distance and the redshift, we can then estimate $H_0$. 

This effect necessarily has a large uncertainty (due to the fact that peculiar velocities are unknown). As a result, the effect is often small compared to the signal provided by the known motion of the Earth.

However, in some cases, this statistical effect is as important as (or even more important than) the proper motion due to actual parallax. For example, in cases where the proper motion due to the Earth's motion is small compared to the uncertainty in proper motion measurements (either because of high uncertainty or large distances), the signal due to Earth's motion is weak. As a result, the signal is dominated by this extragalactic statistical parallax effect, and an increase in the peculiar velocity dispersion actually leads to a decrease in the uncertainty on the measurement of $H_0$. This is discussed further in Section \ref{sec:stat_parallax_results}. 

\section{Results}
\label{sec:results}

\subsection{Cosmological Parameters}
We used the following cosmological parameters in the Fisher information analysis and in the simulation.

The current radiation energy density parameter was set to $\Omega_R = 9 \times 10^{-5}$. The current matter energy density parameter was set to $\Omega_M = 0.31$. The current dark energy density parameter $\Omega_\Lambda$ was set to $1-\Omega_R-\Omega_M$, rendering the Universe flat \citep{ryden,planck}. 

In this section, we analyze a survey of a universe with Hubble parameter equal to $H_0 = 70$ km s$^{-1}$ Mpc$^{-1}$, galaxy peculiar velocity dispersion equal to $\sigma_v = \frac{660}{\sqrt{3}}$ km/s \citep{padilla}, and speed of the solar system relative to the CMB equal to $\langle25.8 \text{ km/s}, 246 \text{ km/s}, 271 \text{ km/s}\rangle$ in standard Galactic coordinates \citep{gordon}.

Throughout this analysis, we hold the fractional uncertainty in measured redshift fixed at $0.001$. Each galaxy is assumed to have the same uncertainty $\sigma_p$ in both components of its proper motion.

\subsection{Trials}
To relate comoving distances to cosmological redshifts, we numerically solved the Friedmann equation at fifty evenly spaced redshift values between $0$ and $1$ and then used these points to interpolate the relationship throughout the simulation.

The Monte Carlo integration in the Fisher analysis was performed with $1\,00000$ points. 

The simulation generated between $1000$ and $10000$ surveys for a given set of experimental parameters (depending on the point in parameter space). From these surveys, the standard deviation of the maximum-likelihood estimates of $H_0$ was taken as the uncertainty in $H_0$ and the mean value of the maximum-likelihood estimates was taken as the expected value of $H_0$.

All numerical optimizations were performed with SciPy's Powell optimization method, and all numerical integrations were performed with SciPy's quadrature integration method.

To ensure that each simulation generated enough surveys for reliable results, we ran the simulation twice for each point in parameter space and compared the percent error in the uncertainty estimates of the two runs. The maximum percent error was $4.92 \%$. The minimum percent error was $2.05 \times 10^{-6} \%$. The mean percent error was $1.30 \%$, and the standard deviation of the percent errors was $1.14 \%$.

\subsection{Fisher Analysis Results}
\label{sec:fisher_results}

\begin{figure}
    \centering
    \includegraphics[width=3.5in]{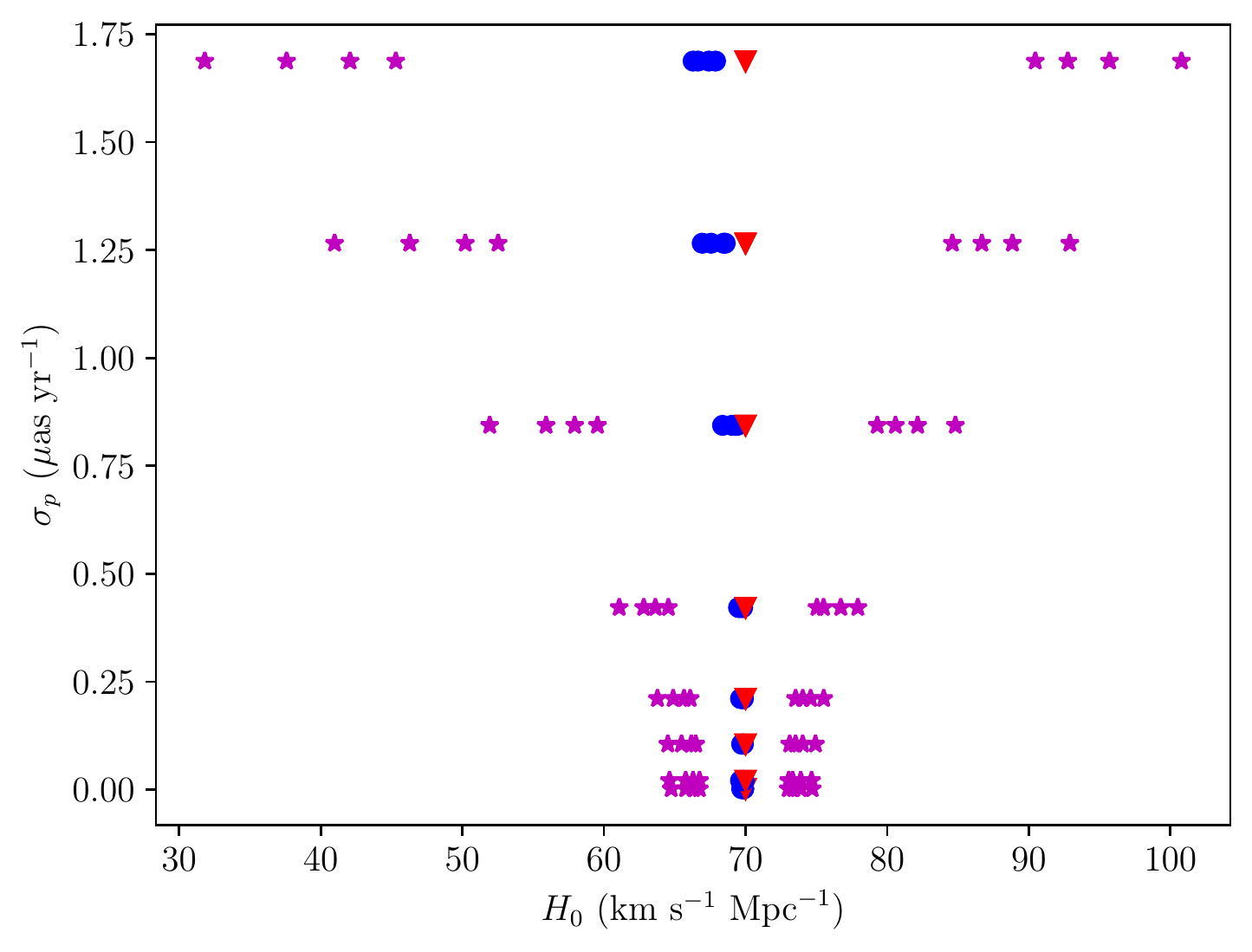}
    \caption{This plot displays the bias for a set of survey parameters with the maximum survey redshift held fixed at $0.058$, where $\sigma_p$ is the proper motion uncertainty. The red triangles mark the true value of $H_0$ in the simulation, $70$ km s$^{-1}$ Mpc$^{-1}$. The blue points mark the expected value of the maximum-likelihood estimator of $H_0$ for the simulation of a given parameter set. The magenta stars mark the mean value minus the standard deviation and the mean value plus the standard deviation. The different points for the same proper motion uncertainty correspond to surveys with different numbers of galaxies ranging from $50$ to $125$.}
    \label{fig:bias_plot}
\end{figure}

As mentioned in Section \ref{sec:fisher}, the Cram\'er-Rao Inequality requires an unbiased estimator to be guaranteed. Figure \ref{fig:bias_plot} shows that the maximum-likelihood estimator has a slight negative bias, which is always small compared to the uncertainty. The ratio of the bias in estimated $H_0$ to the error in $H_0$ ranges from $-0.03$ to $-0.2$, with a median of $-0.08$. However, this ratio is smaller for low $\Delta H_0$ values than for higher ones, typically by a factor of roughly $2$. The bias is therefore relatively small for the results of interest.

A possible explanation for this bias is the volume element in our prior and our assumption about uniform volume density of galaxies. All else being equal, our model prefers larger distances for galaxies. This is simply because there is more volume at larger distances - if we know nothing else about a galaxy, it is more likely to be found at larger distances because of the higher volume and uniform density. The estimator therefore tends to select the largest distance compatible with the redshift observation. For a fixed redshift, increasing the distance corresponds to a reduced $H_0$, producing the effect we observe. While this is a reasonable explanation for the effect, one should also bear in mind that an estimator has no guarantee of being unbiased. An exact explanation of the source of the bias does not necessarily exist.

As discussed in Section \ref{sec:fisher}, the Cram\'er-Rao lower bound for a biased estimator is well approximated by the reciprocal of the Fisher information when the derivative of the bias with respect to the estimator is low. This is typically the case for our estimator; for example, we numerically estimated the derivative of the bias at $H_0 = 70$ km s$^{-1}$ Mpc$^{-1}$ with $\sigma_p = 0.211$ $\mu$as yr$^{-1}$, $N=216$, and a maximum survey redshift of $0.070$, finding it to be approximately $b'(H_0) = 0.0088$. Since $b'(H_0) \ll 1$, the correct value for the Cram\'er-Rao lower bound is very close to the reciprocal of the Fisher information. We therefore use the unbiased version of the Cram\'er-Rao inequality to quantify the standard deviation of our estimator, taking the reciprocal of the square root of the Fisher information to be the estimate of the uncertainty in $H_0$.

For each point that we simulated in parameter space, we estimated the uncertainty in $H_0$ using the Fisher information. The set of ratios of simulation uncertainty estimation to Fisher uncertainty estimation (as given by equation (\ref{eq:cramer-rao})) had a minimum of $0.926$, a maximum of $1.151$, a mean of $1.042$, and a standard deviation of $0.029$. 
In summary, the uncertainty in the maximum-likelihood estimator is always extremely close to the Cram\'er-Rao bound.

\begin{figure*}
    \centering
    \fbox{\includegraphics[width=6in]{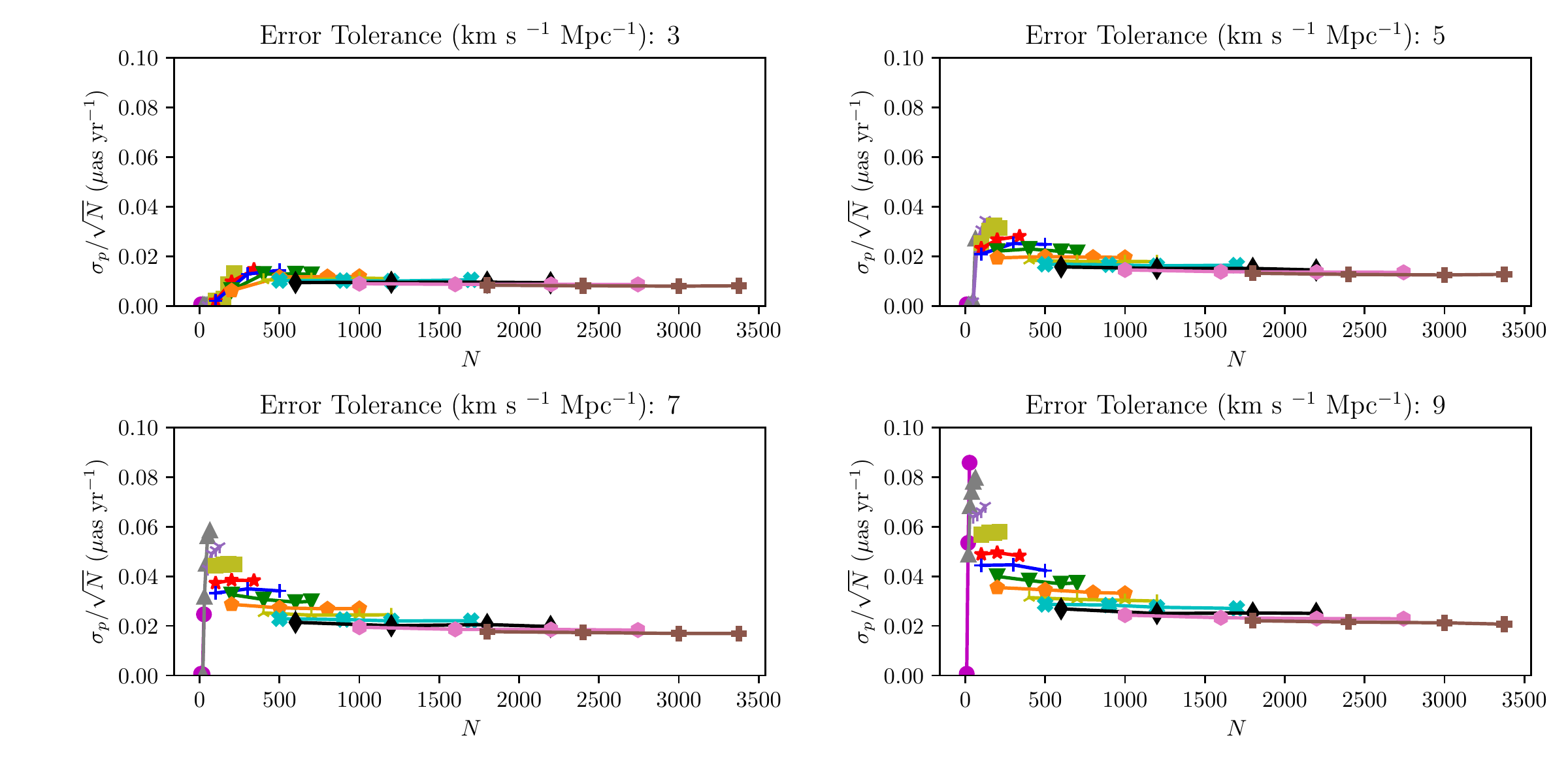}}
    \caption{The figure of merit is graphed versus number of galaxies for a specified error tolerance in $H_0$, where $\sigma_p$ is the proper motion uncertainty and $N$ is the number of galaxies surveyed. The different colors of lines correspond to different maximum survey redshifts as follows: 0.035 (magenta), 0.046 (gray), 0.058 (purple), 0.070 (olive), 0.081 (red), 0.093 (blue), 0.104 (green), 0.116 (orange), 0.128 (yellow), 0.139 (cyan), 0.151 (black), 0.162 (pink), 0.174 (brown). For each maximum redshift, the maximum value of $N$ is determined by the requirement that galaxy separations exceed the velocity correlation length.}
    \label{fig:uncorrelated_fourpanel}
\end{figure*}

\subsection{Simulation Results}
\label{sec:simulation_results}
We imagine a survey with a fixed number of galaxies, a maximum survey redshift, and desired error tolerance in the determination of $H_0$. We then use the results of the simulation to interpolate the necessary precision of proper motion measurement ($\sigma_p$) required to meet this error tolerance.

For each redshift limit, we limit the number of galaxies such that the average distance between galaxies in the simulation is greater than or equal to the velocity coherence length, which we take to be $50$ Mpc. 

We chose this value based on data from the Galacticus simulation from the CosmoSim suite of simulations \citep{Knebe_2017, BENSON2012175}. Peculiar-velocity correlations in these simulations appear to be broadly consistent with observational data \citep{wang}. For our purposes, a simulation-based estimate is cleaner than an observational result, as we are interested in knowing the correlation function of arbitrary components of the 3D peculiar velocity field. Due to both cosmic variance and the fact that only radial components can typically be measured, the relationship between this quantity and observational data is complicated. In any case, for present purposes a reasonable estimate of the coherence length is all that is required.

To be specific, we computed the two-point velocity correlation function for each component of galaxy velocities for a cubic neighborhood of CosmoSim data and found the distance by which the two-point correlation function had dropped by approximately one $e$-fold. This cubic neighborhood had a proper side length of $150$ Mpc and roughly $35000$ galaxies of mass at least  $10^{12}M_\odot$ within it.

Figure \ref{fig:uncorrelated_fourpanel} displays the required sensitivity in term of a figure of merit, 
\begin{equation}
\mathrm{FOM} =\frac{\sigma_p}{N^{1/2}},
\end{equation} 
where $\sigma_p$ is the proper motion uncertainty and $N$ is the number of galaxies surveyed. Under the assumption that the precision of a proper motion measurement improves in proportion to the square root of the observing time, this FOM is proportional to $T^{-1/2}$, where $T$ is the total observing time devoted to the survey.\footnote{Note that $T$ is the telescope time, not the total elapsed time, for the survey. Proper motion measurements will improve in direct proportion to the total elapsed time. Our figure of merit assumes that the latter is fixed.} Larger FOMs therefore correspond to ``easier" surveys.

\subsection{Velocity Field Reconstruction Results}
\label{sec:correlation_results}
As described in Section \ref{sec:velocityreconstruction}, we compared the results of the uncorrelated survey to the results of the survey with velocity field reconstructions. For the subset of parameter space with redshift limits of $0.058$ and below, we compute the uncertainty in $H_0$ via simulations in which $0.25, 0.5$, and $0.75$ of the variance of peculiar velocity components is due to this known and correlated velocity field. The $\sigma_v^2$ used in the likelihood function is therefore the fraction of the variance attributable to the random component of the velocity. 

We found the ratio of the error estimate of the uncorrelated simulation to the error estimate of the simulation with $0.75$ of the velocity having been reconstructed. We then computed the maximum, minimum, mean, and standard deviation of these ratios. The results are found in Table \ref{table}.
  
\begin{table}
    \caption{Ratios of error estimates based on the assumption of uncorrelated errors to those based on the assumption that 75\% of the peculiar velocity variance is correlated and has been reconstructed, as described in Section \ref{sec:velocityreconstruction}. }
    \label{table}
    \centering
\begin{tabular}{|c|c|c|c|c|}
     Max Redshift & Min & Max & Mean & Standard  \\
     Redshift & Ratio & Ratio & 
     Ratio & Deviation\\
     \hline
     0.035 & 1.15 & 1.59 & 1.38 & 0.16 \\
     0.046 & 1.17 & 1.60 & 1.39 & 0.14 \\
     0.058 & 1.17 & 1.62 & 1.40 & 0.14 \\
     \hline
\end{tabular}
\vskip 0.2in
\end{table}

\begin{figure*}
    \centering
    \fbox{\includegraphics[width=6in]{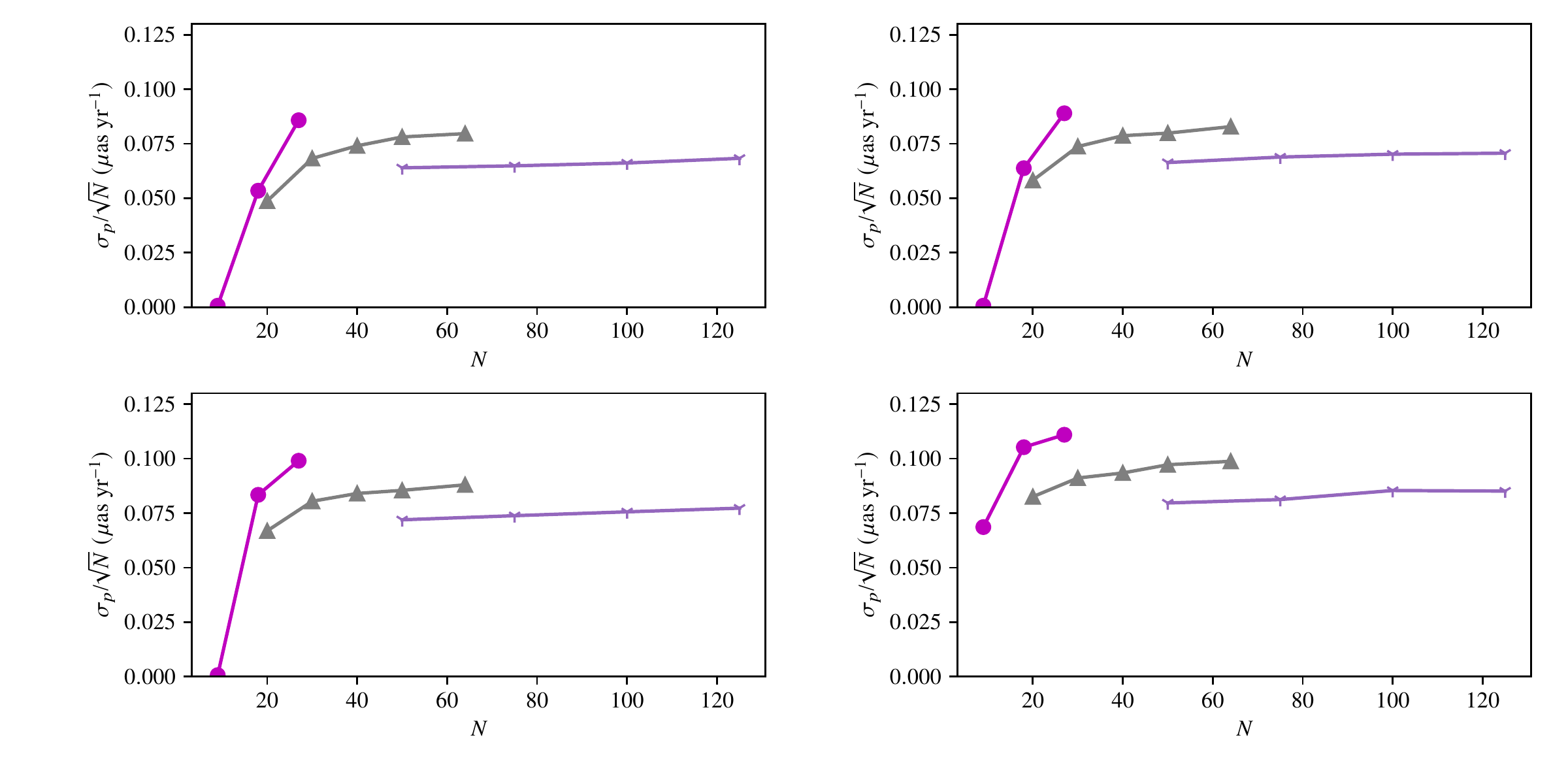}}
    \caption{The figure of merit is graphed versus number of galaxies for an error tolerance in $H_0$ of $9$ km s$^{-1}$ Mpc$^{-1}$, where $\sigma_p$ is the proper motion uncertainty and $N$ is the number of galaxies surveyed. The upper left panel has $0$ of the peculiar velocity reconstructed, the upper right panel has $0.25$ of the peculiar velocity reconstructed, the lower left panel has $0.5$ of the peculiar velocity reconstructed, and the lower right panel has $0.75$ of the peculiar velocity reconstructed. The colors are as in Figure \ref{fig:uncorrelated_fourpanel}.}
    \label{fig:reconstructed_fourpanel}
\end{figure*}

Figure \ref{fig:reconstructed_fourpanel} shows FOM plots under the assumptions that the fraction of peculiar velocities that are correlated and reconstructed is 0.25, 0.50, and 0.75.

The differences resulting from peculiar velocity reconstruction are small, providing evidence that peculiar velocity correlations do not dramatically affect our results. 

In addition to the primary tests described above, we performed additional tests as described in Section \ref{sec:velocityreconstruction}. We now detail the results of these additional tests.

Given the peak in Figure \ref{fig:uncorrelated_fourpanel}, we are primarily interested in the lower redshift surveys. We therefore analyzed the worst-case scenario (of correlated velocities with correlation completely ignored in the fitting of $H_0$) for the same survey sizes and proper motion uncertainties as the original data set for the lower redshift surveys. For the subset of the parameter space which we analyzed in this worst case scenario, the ratio of $\Delta H_0$ in the worst case scenario to $\Delta H_0$ in the original analysis had a mean of $1.85$, a minimum of $1.04$, a maximum of $3.41$, and a standard deviation of $0.60$.

\begin{figure*}
    \centering
    \includegraphics[width=6in]{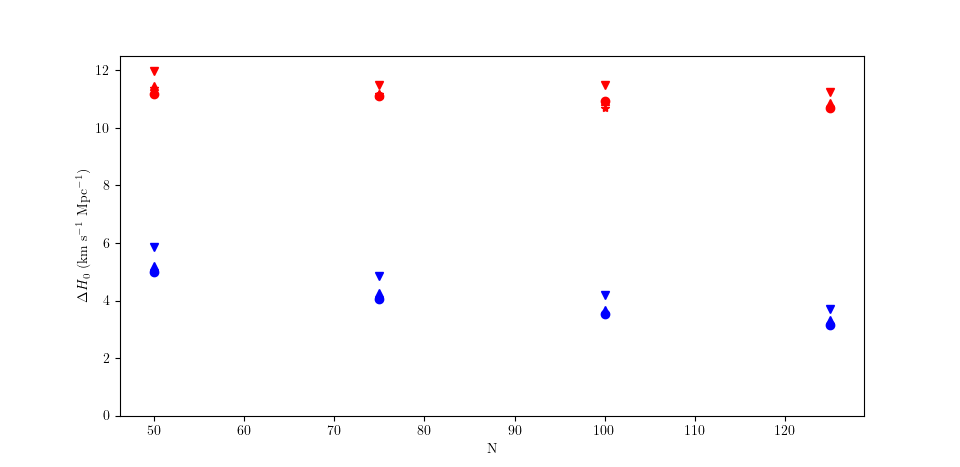}
    \caption{The uncertainty in $H_0$ is graphed versus number of galaxies for various values of $\sigma_p$ (corresponding to the different curves). The blue curves correspond to the original analysis; the red curves correspond to the worst case scenario analysis. Each $\sigma_p$ value is associated with a specific indicator, so the red and blue curves with the same indicators are for the same $\sigma_p$ values.}
    \label{fig:comparison_slice}
\end{figure*}

\begin{figure*}
    \centering
    \fbox{\includegraphics[width=2.8in]{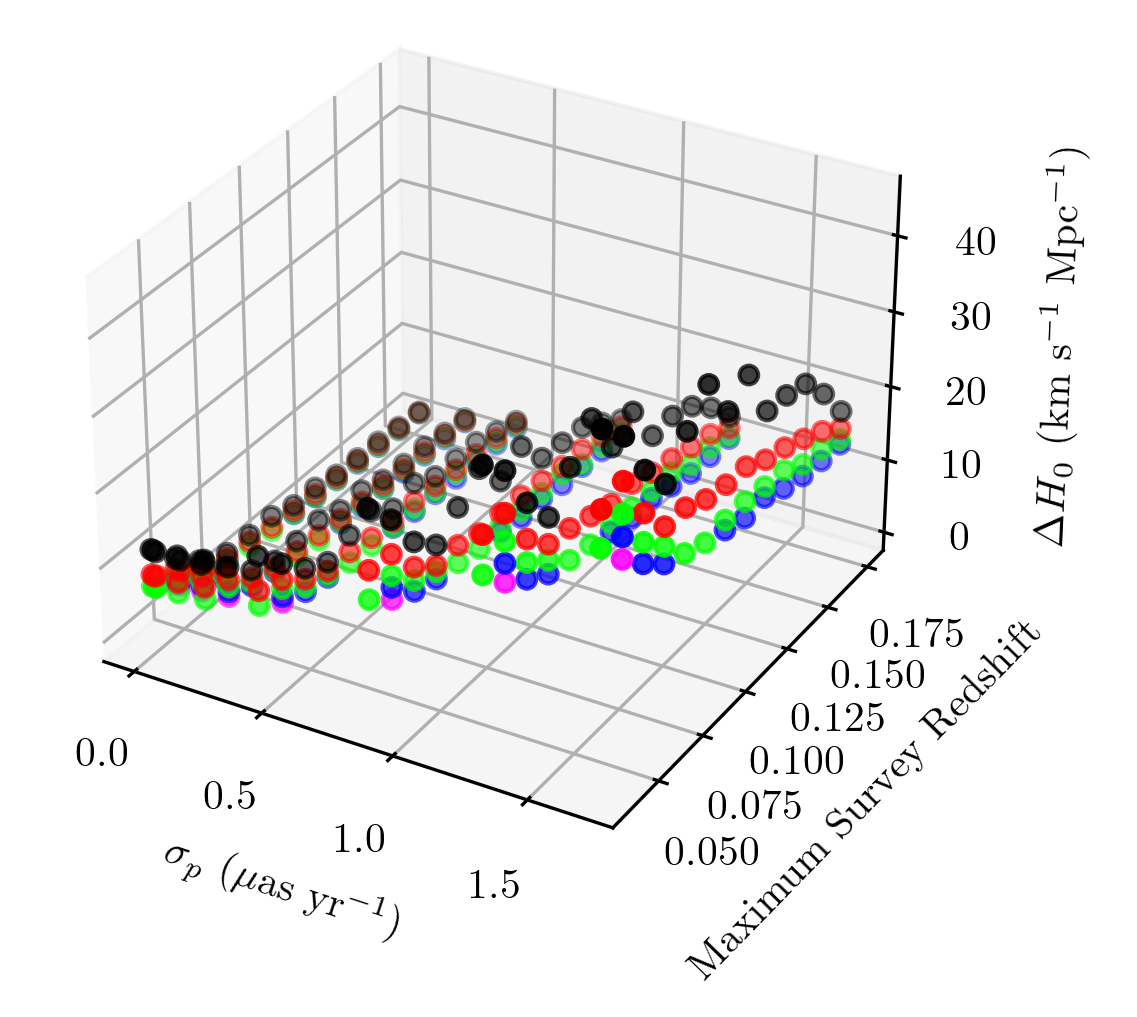}\hskip 0.5in
    \includegraphics[width=2.8in]{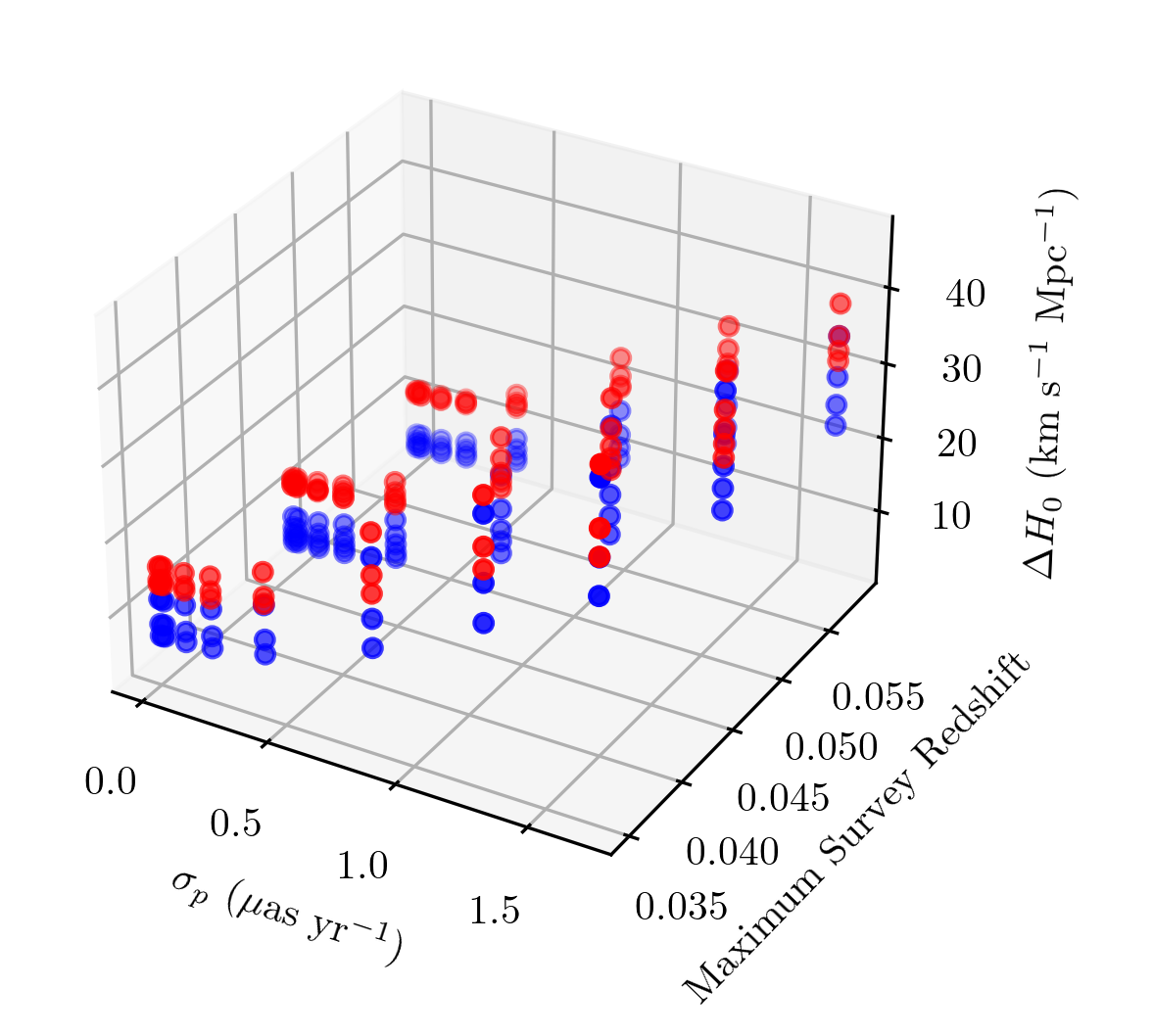}
    }
    \caption{ The uncertainty $\Delta H_0$ in the measurement of the Hubble constant is plotted as a function of the survey parameters. The proper motion uncertainty $\sigma_p$ and the maximum survey redshift are displayed on the axes. In the left plot, different survey sizes for these parameters are indicated by different colored dots. These results are for the entirety of parameter space in the original analysis. In the right plot, the blue points correspond to the original analysis while the red points correspond to the worst case analysis. These results are for the subset of parameter space analyzed via both methods. The multiple points occurring at any given $\sigma_p$ and maximum redshift correspond to different survey sizes.}
    \label{fig:multidim_twofigure}
\end{figure*}

Figures \ref{fig:comparison_slice} and \ref{fig:multidim_twofigure} display $H_0$ error estimates as a function of parameter space for both our original and worst case analysis. Figure \ref{fig:comparison_slice} shows these estimates for a slice of parameter space, while \ref{fig:multidim_twofigure} shows these estimates for all points in parameter space which were analyzed using both methods. The comparison of these results shows that while our approximation of uncorrelated velocities is not a negligible error, it still gives us both the correct shape of the parameter space and estimates of the $H_0$ uncertainty that are typically within a factor of two. These similarities support our claim that our approximation is good enough to draw useful conclusions about the order of magnitude of errors involved in a future parallax survey and about the optimal design of such a survey.

\subsection{Extragalactic Statistical Parallax Results}
\label{sec:stat_parallax_results}
\begin{figure*}
    \centering
    \fbox{\includegraphics[width=6in]{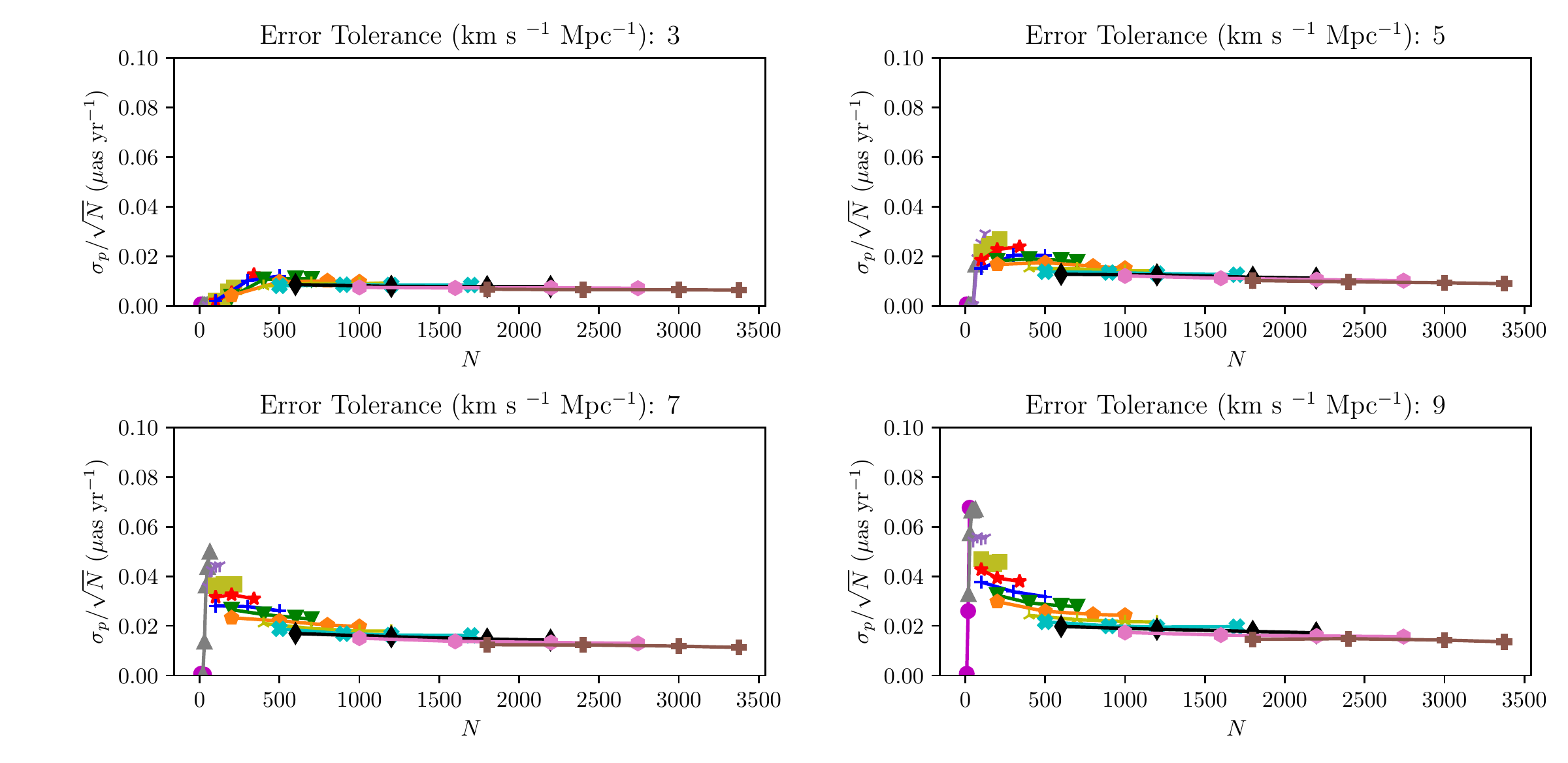}}
    \caption{The figure of merit is graphed versus number of galaxies for a specified error tolerance in $H_0$, where $\sigma_p$ is the proper motion uncertainty and $N$ is the number of galaxies surveyed. In contrast to Figure \ref{fig:uncorrelated_fourpanel}, our Galaxy's peculiar velocity is set to zero, so that only the effects of extragalactic statistical parallax are included. The colors are as in Figure \ref{fig:uncorrelated_fourpanel}.}
    \label{fig:statistical_fourpanel}
\end{figure*}

To illustrate the information obtainable from ``statistical parallax,'' we performed simulations in which the Earth's peculiar velocity was set to zero, so that there is no overall secular parallax. The results are shown in Figure \ref{fig:statistical_fourpanel}.

Comparison of Figure \ref{fig:statistical_fourpanel} to Figure \ref{fig:uncorrelated_fourpanel} shows that in the simulation with uncorrelated velocities, a significant fraction of the signal is due to statistical parallax. However, this comparison presumably overestimates the effect of statistical parallax in relation to actual secular parallax. First, the assumption that peculiar velocities are completely unknown causes an underestimation of the signal due to well-known motions. See Figure \ref{fig:reconstructed_fourpanel} and note that as the fraction of reconstructed peculiar velocity increases, the uncertainty in $H_0$ decreases. Second, the assumption that peculiar velocities are uncorrelated most likely causes an overestimation of the signal due to statistical parallax. Since all of the information in statistical parallax is due to knowledge of the statistical distribution of peculiar velocities, correlating the velocities presumably results in less information from the same number of galaxies.

Unfortunately, statistical parallax in the actual Universe would be significant in the case where velocity correlations are important but unknown. This scenario is not well approximated by either of the simulations: the first simulation has uncorrelated peculiar velocities, while the second simulation has known correlations. We therefore cannot test a more realistic case of statistical parallax using this model, so we cannot test the hypothesis that uncorrelated peculiar velocities leads to an overestimate of the usefulness of statistical parallax.

\section{Discussion}
The plots in Figure \ref{fig:uncorrelated_fourpanel} have the same general features, several of which are to be expected. For each error tolerance, the FOM has a clear peak at a maximum redshift between 0.045 and 0.058, caused by competing effects. If the redshift limit is small, there is not enough volume to survey a large number of galaxies while maintaining the constraint that galaxy separation exceed the correlation length. (This is the reason that the low-redshift curves stop at small values of $N$.) At high redshift, on the other hand, the parallax signal becomes small and hard to measure.

A useful observation for experimental design is the relative scale of the curve and the steepness of the two sides. While there is a clear peak in the figure of merit, it is within a factor of roughly four of the figure of merit for the higher redshift limits that we tested -- that is, the peak FOM is of the same order of magnitude as those with greater redshift limits. Furthermore, the curve is much steeper to the left of the peak than to the right. This indicates that if a survey design is to deviate from the ideal redshift limit, it is better for the redshift limit to be higher than the ideal redshift limit than for it to be lower.

Although our formalism does not allow us to forecast errors for surveys in which galaxies are closer together than the velocity correlation length, there is reason to believe that the error forecasts for surveys with larger numbers of more closely-spaced galaxies would have similar FOMs -- that is, that the curves in Figure \ref{fig:uncorrelated_fourpanel} and the following figures would roughly plateau rather than decline dramatically if extended to higher $N$ at a fixed redshift cutoff. In particular, in the limit where the separation between a set of galaxies is much less than the velocity correlation length, we could treat the  peculiar velocities as identical and treat the entire collection as a single object with a correspondingly larger observing time. In this limit, the FOM is independent of the number of galaxies in that collection. To forecast errors for a survey with mean galaxy separation less than the correlation length, we could imagine dividing the survey volume into voxels whose size is of order the correlation length and treating all galaxies in each voxel as a single data point in this manner. (Of course, one would not analyze the real data in this way, but it is plausible that such an approach would give decent enough error forecasts for present purposes.) This approximation would lead to plateaus in the various curves -- that is, one would achieve results similar to the optimum FOM even if one shifted to larger values of $N$ at a given redshift limit.

Finally, we wish to compare the precision of measurement required for a useful constraint of $H_0$ with the capabilities of present and near-future experiments.
For example, with an error tolerance of $7$ km\,s$^{-1}$\,Mpc$^{-1}$, the optimum FOM is $\sim 0.06\,\mu\mathrm{as\,yr}^{-1}$, obtained by surveying $\sim 64$ galaxies with redshift limit $\sim 0.046$. This corresponds to a precision for each measurement of $\sigma_p\sim 0.5\,\mu\mathrm{as\,yr}^{-1}$.
While this is certainly a greater degree of precision than is currently available, it is not vastly beyond the capabilities of current astrometric surveys. For example, \citet{Gaia_parallax} forecast a Gaia end-of-mission catalog with $10^4$ nearby galaxies with mean proper motion uncertainties of about 70 $\mu\mathrm{as\,yr}^{-1}$, leading to an FOM of about $0.7\,\mu\mathrm{as\,yr}^{-1}$, comparable to our optimal forecast. In the future, instruments such as the Nancy Grace Roman Space Telescope \citep{roman}, the Vera C. Rubin Observatory \citep{rubin}, or the next-generation VLA \citep{ngvla} may provide the capability to perform even better surveys.

As we noted in the introduction, the calculations presented in this paper are meant as preliminary estimates. Although they are based on approximations that limit the precision of the error forecasts, we have presented tests that show that the resulting errors are not so large as to prevent our results from providing a useful guide at this early stage of consideration of this method. Naturally, as details of a survey design come into focus, more precise error forecasts will be required.

The methods we have presented here can be extended in various ways to assist in designing the optimum design for a future proper motion survey. For example, we have assumed an all-sky redshift-limited survey, but our simulations can easily be generalized to accommodate partial sky coverage or more complicated redshift and/or brightness selection functions. In future work, we plan to consider whether including the strongly lensed galaxies that are expected to be found in future surveys \citep{oguri} can enhance the precision of the measurement, by increasing the size of the proper motion of some images and/or by using multiple images to reduce the uncertainty.

\section*{Acknowledgements}
This work was supported by an Undergraduate Science Research Fellowship grant from the Virginia Foundation of Independent Colleges as well as by the University of Richmond.
The CosmoSim database used in this paper is a service by the Leibniz-Institute for Astrophysics Potsdam (AIP).

\section*{Data Availability}

No new observational data were obtained as part of this work. Code used in the computations is available upon request to the authors.

\bibliography{bibliography}

\bsp	
\label{lastpage}
\end{document}